# Are there missing bond paths in Trimethylenemethane-Iron-tri-carbonyl,(CO)$_3$Fe-(C$_4$H$_6$), complex?

Comment on: "Bonding Analysis of Trimethylenemethane (TMM) Complexes [(CO)$_3$M-TMM] (M = Fe, Ru, Os, Rh$^+$). Absence of Expected Bond Paths [J. Organomet. Chem. (2013) doi: 10.1016 /j.jorganchem. 2013.03.047]"

Shant Shahbazian

Faculty of Chemistry, Shahid Beheshti University, G. C., Evin, Tehran, Iran, 19839, P.O. Box 19395-4716.

Tel/Fax: 98-21-22431661

E-mail:
(Shant Shahbazian) chemist_shant@yahoo.com

# Abstract

In a recent paper [J. Organomet. Chem. (2013) doi: 10.1016 /j.jorganchem. 2013.03.047] analyzing the bonding mode of Trimethylenemethane (TMM) with some metal carbonyls, Mousavi and Frenking have declared the absence of bond paths, the so-called missed bond paths, between metal atoms and terminal carbon atoms in several complexes. They inferred these missed bond paths based on two principles, first, the fact that both the molecular orbital diagrams and the energy decomposition analysis point to the presence of chemical bonds between the metal atoms and the terminal carbon atoms and second, the presupposition that the indicator of a chemical bond within the quantum theory of atoms in molecules (QTAIM) is a bond path. They used these observations and concomitant interpretation to question the reliability of bonds paths as indicators of chemical bonds. In this communication, it is first demonstrated that the presupposition of the equivalence of a bond path and a chemical bond within the context of the QTAIM is superficial and basically flawed which is not only against with the recent strict declaration on the contrary [R.F.W. Bader, J. Phys. Chem. A 113 (2009) 10396], but also in odd with the foundations of the QTAIM. Then, it is demonstrated that the so-called missed bond paths indeed appear in molecular graphs of some non-equilibrium geometries that are energetically quite accessible at room temperatures. To emphasize on the importance of this observation the term *passionate neighbors* is coined referring to the atomic basins that do not share an inter-atomic surface at the equilibrium geometry but are neighbors, share an inter-atomic surface, at non-equilibrium geometries accessible by nuclear vibrations. Using the delocalization index as well as other evidence from previous literature it is demonstrated that the QTAIM analysis is indeed compatible with the presence of chemical bonds between iron metal and terminal carbons in $(CO)_3$Fe-TMM complex. This observation further demonstrates that a consistent bonding pattern is emerged from a combined careful QTAIM analysis and the other theoretical approaches used for the analysis of bonding modes in aforementioned complexes. Finally, some general discussions are done to unravel the delicate relationship between the QTAIM proposed concepts, e.g. bond paths and molecular graphs, and some orthodox concepts of chemistry, e.g. chemical bonds and chemical structures, emphasizing that there is no one-to-one relationship between these two categorizes.

# 1. Introduction

Recently, in a computational paper [1], Mousavi and Frenking have analyzed the bonding of Trimethylenemethane (TMM) with metal carbonyls in some complexes: [(CO)$_3$M-TMM] (M= Fe, Ru, Os, Rh$^+$). Apart from considering the usual qualitative Molecular Orbital (MO) correlation diagrams as well as the Energy Decomposition Analysis (EDA), the authors have also used the results of what they called atoms in molecules analysis unraveling the bonding in these complexes. From their MO and EDA analysis, they concluded that the central metals are chemically bonded to both the terminal and central carbons atoms of TMM then proceeded the topological analysis of the one-electron density of the complexes extracting the molecular graphs (MG) as well as considering the Laplacian maps of the computed one-electron densities. The reported MG, computed just at the equilibrium geometries of the all considered complexes, did not contain bond paths (BP) between the metal atoms and the terminal carbons while a BP was observed between the metal atoms and the central carbons. From this observation the authors came to the conclusion that: "… there are bond paths between the metal atom and the central carbon atom but not to the terminal carbon atom which comes from the different interatomic distances. This clearly shows that the absence and the existence of a bond path are not reliable criteria for chemical bonding". Strangely, without explicitly crediting, this is exactly the duplication of the observation and interpretation announced first by Farrugia and coworkers in 2006 in a paper titled: "Chemical Bonds without "Chemical Bonding"? A Combined Experimental and Theoretical Charge Density Study on an Iron Trimethylenemethane Complex" (though the authors have cited this paper in their reference list (ref. 23), it has been cited just as a source of the experimental geometry of (CO)$_3$Fe-TMM). Seemingly, like Farrugia and coworkers, the authors of the original paper have tacitly assumed that a BP is an indicator of chemical bonding within the context of the quantum theory of atoms in molecules (QTAIM) [3] so came to the conclusion that their computational study reveals evidence that discredits the reliability of such bonding indictor as is also highlighted in part of the title of their paper: "absence of expected bond paths".

Oddly enough, the authors did not cite the seminal paper of Richard Bader entitled [4]: "Bond Paths Are Not Chemical Bonds" in response to similar claims made on the equivalence of BP and chemical bonds, basing their whole reasoning on a flawed assumption regarding the relation of the QTAIM analysis and chemical bonds. Although the paper of Bader was appeared at 2009, even since then one occasionally finds papers with similar claims announcing discovery of MG that discredit the equivalence of BP and chemical bonds; the paper of Mousavi and Frenking belongs to this group of papers. As is discussed in the rest of this communication, even neglecting the illuminating paper of Bader, the mathematical structure of the QTAIM unveils clearly that not only BP are not generally indicators of chemical bonds but also MG are not generally the orthodox chemical structures thus undermining the superficial interpretations of the QTAIM analysis.

## 2. The QTAIM and Chemical Bond

### 2.1. Bypassing confusing terminology

In the above mentioned paper Bader emphasizes from the outset on the role of "Bad Grammar" on triggering controversies about the meaning of BP [4]. He insists that the word *bond* or as chemists usually used *chemical bond* is intrinsically a floppy concept that lacks a clear cut definition (*vide infra*) so he proposes the word *bonded* to be used/replaced that indicates the presence of BP between two atoms. Accordingly, in this terminology, the statement "two atoms are bonded" is *not* equivalent to the statement "there is a chemical bond between two atoms". The former statement is meaningful only within the context of the QTAIM while the latter is a general statement within the framework of the fuzzy language of chemistry. In other words, "QTAIM-bonded" atoms are identified if and only if a QTAIM analysis reveals BP between two atoms thus unveiling neighboring atomic basins that share a common inter-atomic surface (IAS) [3]. This is easy to comprehend, remembering that in both helium dimmer and nitrogen molecule, as two extreme examples from the energetic viewpoint, the atoms are linked by BP; while all chemists agree that there is chemical bond between nitrogen atoms in nitrogen molecule, they prefer to describe the helium dimmer as a van der Waals complex with no chemical bond. A more thorough computational analysis of some van der Waals complexes indeed reveals that many BP appear between atoms in these complexes that barely any chemist inclines to assume a chemical bond in between [5]. Accordingly, the presence of BP between atoms that are not usually assumed to have chemical bond in between is a common feature of MG of many large crowded molecules as well as molecular aggregations in crystals which demonstrates that there is *no one-to-one equivalence between BP and chemical bonds* [6]. As Bader hoped, it is probable that in long period the fuzzy concept of chemical bond to be dismissed in favor of the rigorous concept of a BP however in the meantime, leaving aside speculations on the future of chemistry, these two concepts coexist, and the distinction between these two terms is of crucial importance. Indeed, the words bonded/bonding within the context of the QTAIM are *not* synonyms for chemical bond so in the rest of this communication, to prevent any ambiguity/misunderstanding, the QTAIM-bonded atoms are just called *neighboring* atoms. Even more, recognizing that chemical bond is not redefined within context of the QTAIM is important, thus *the QTAIM is not a theory of chemical bond and has no unique index/indictor of chemical bond*. Accordingly, the relation of the QTAIM central concepts with the concept of chemical bond is more intricate than is usually conceived as is emphasized in subsequent section.

At present stage, based on this background, it suffices to conclude that in general, the absence of BP between two atoms that most chemists incline to assume a chemical bond in between, in contrast to the conclusion of the authors, is not a sign of the weakness of the QTAIM because its mission is *not* recognizing chemical bonds. However, as is discussed in subsequent section, there are subtle points regarding the so-called absent or missed BP that do not seem to be noticed by the authors of the original paper. To sum up this subsection, declaring the "absence of expected bond paths" just stems from a misunderstanding that within the context of the QTAIM a one-to-one equivalence between BP and chemical bonds is expected; when decoupling these two concepts, no reason remains for the "expected bond paths". *BP are just emerging from the topological analysis of the one-electron density while chemical intuition is not a safe way, particularly in large and crowded molecules, to predict their presence and location.*

2.2. MG: Beyond equilibrium geometries

In most computational QTAIM studies there is a tendency to consider the MG and associated BP at equilibrium geometries. However, molecules are not static objects and the eternal nuclear vibrations guarantee that apart from equilibrium geometry many other nuclear configurations are also accessible for a molecule. Rigorously specking, a probability density corresponds to each nuclear configuration emerging through solving the Schrödinger equation for nuclear vibrations. While, based on harmonic oscillator model of vibrations it is usually assumed that the equilibrium geometry is the most probable among the other geometries, since not all vibrational normal modes are stiff, many non-equilibrium geometries are also accessible as well (assuming non-harmonic dynamics for nuclear vibrations, even it is not possible to claim that the equilibrium geometry is the most probable geometry). Although this *dynamic* aspect of molecules is neglected in most bonding analysis, it may have a pivotal role in the topological analysis of the QTAIM as is demonstrated in the rest of this subsection.

For any molecular geometry, equilibrium or non-equilibrium, the topological analysis of the one-electron density yields a MG and the corresponding BP (for non-equilibrium geometries, instead of BP, the phrase *atomic interaction lines* (AIL) is used [3]). As has been comprehensively disclosed in the original literature of the QTAIM [3], different molecular geometries may have different MG and for a systematic survey geometries must be categorized into sets in a way that a single MG is attributed to each set. Since nuclear vibrations are always at work, even at absolute zero temperature, the question naturally emerges whether it is legitimate to attribute a single MG to a molecule. For a limited number of simple molecules, systematic survey of more accessible geometries demonstrates that indeed the MG of the equilibrium and neighboring non-equilibrium geometries is the same or in other words they all belong to the same set/class of topological structures [3]. However, this is not a general rule and although systematic studies on MG of non-equilibrium geometries are rare, at least for certain species it was demonstrated that various MG are easily accessible during nuclear vibrations [7,8] and there is no reason to insist that the *one-molecule one-MG* is a universal paradigm. Accordingly, attributing a single MG to a molecule is just a simplification/approximation that is acceptable for cases where single MG dominants a large set of more accessible geometries. In practice it is hard to estimate the relative importance of various MG of a molecule, which needs a large sampling of non-equilibrium geometries then unraveling the corresponding MG and associated probability of occurrence. However, relegating computational demands, there is no doubt that in the case of competing/coexisting MG, to have a complete picture of the QTAIM analysis, one can not just concentrate on a single MG even if the corresponding set contains the equilibrium geometry. As is demonstrated in rest of this subsection (also emphasized briefly by Farrugia and coworkers [2]), considering $(CO)_3$Fe-TMM complex as an illustrative example, the cases considered by Mousavi and Frenking also belong to the class of *many-MG* thus their conclusion that was based on analyzing just the MG of equilibrium geometry is deceptive.

In order to consider the MG of $(CO)_3$Fe-TMM complex, this molecule was first optimized at BP86/6-311+g(d,p) computational level, similar to the level employed in the original study [1], without imposing any symmetry constraint and then the harmonic frequencies of vibrational normal modes were derived. Comparing the optimized structure with Figure 1 of the original paper as well as Figure 1 of the paper of Farrugia and coworkers [2], indeed reveals that the same reported geometry is reproduced. In this account, besides the topological analysis of the one-electron density of the optimized geometry, some non-equilibrium geometries following two soft normal modes with wave numbers ~341 and ~ 379 $cm^{-1}$ are analyzed employing the AIM2000 package [9-11] (for the $C_{3v}$ symmetry constrained geometry the first mode is a double degenerate mode labeled E while the second symmetric mode is a non-degenerate mode labeled A1). Figure 1 compares MG of four selected *snapshots* of the 379 $cm^{-1}$ symmetric vibrational motion that all three terminal carbons come near to the iron atom simultaneously whereas Figure 2 similarly compares four MG corresponding to the 341 $cm^{-1}$ normal mode that just a single terminal carbon comes near to iron atom. It is evident from these figures that while the MG of the equilibrium geometry does not have a BP between Fe and the terminal carbons, AIL are appeared between these atoms during the both vibrational modes (Figure 6 in the paper of Farrugia and coworkers [2] depicts a MG of an non-equilibrium geometry similar to that in Figures 1c and 1d). The energies of the non-equilibrium geometries that the Fe-$C_{terminal}$ linking AIL first appeared, Figures 1c and 2c, are ~35.0 and ~1.4 Kcal.$mol^{-1}$ above the equilibrium geometry, respectively, and the new MG remain intact for higher energy structures following these normal modes; since the zero-point energy of the 341 $cm^{-1}$ normal mode is ~1.0 Kcal.$mol^{-1}$, at room temperature, in contrast to high energy structures in Figures 1c and 1d, the novel MG, depicted in Figures 2c and 2d, is quite accessible. Of course the whole used procedure, following just a single normal mode, is an idealization/simplification and in reality the vibrations of a molecule are combinations of various ground as well as low lying excited normal modes thus a far larger set of non-equilibrium geometries are accessible for search on MG containing the so-called "absent" AIL. Accordingly, a brief survey on non-equilibrium geometries, far from a comprehensive study, indeed reveals MG similar to Figures 2c and 2d for geometries that are just ~1.0 Kcal.mol$^{-1}$ higher in energy than the equilibrium geometry. A closer look at Figure 3 of the original paper also reveals that what prevents the presence of a BP between iron and terminal carbons at equilibrium geometry is a slight *penetration* of the basin of central carbon between the mentioned atomic basins. Nevertheless, by a small shift of the equilibrium geometry, the inter-nuclear distance of iron and one of terminal carbons decreases slightly and an AIL appears in between these atoms. Since Figure 3 of the original paper reveals the same penetration pattern for all other (CO)$_3$M-TMM complexes, the above performed analysis on (CO)$_3$Fe-TMM complex and associated conclusion are easily extendable to these complexes as well. To sum up this subsection, the complexes considered by Mousavi and Frenking belong to *one-molecule many-MG* paradigm thus their so-called "expected BP" are appeared in MG of non-equilibrium geometries quite near to the equilibrium geometry. Based on our analysis as well as Figure 3 of the original paper, it seems safe to claim that metal atoms and terminal carbons are *passionate neighboring* atomic basins or in short *passionate neighbors*. Passionate neighbors are pairs of atomic basins that a narrow, third, atomic basin has already penetrated between them and does not let them to have a common IAS (forming a BP) but small changes in molecular geometry can provide the opportunity for these neighbors to share a common IAS and form an AIL.

2.3. Searching for chemical bonds

The identification of chemical bonds is a delicate task that is usually done with a *synthesis* of various geometric, spectroscopic as well as reactivity data derived from experimental and/or theoretical studies best illustrated in the Pauling's "the nature of chemical bond" [12]. Behind this synthesis, it seems there is a basic idea namely, atoms are not interacting with each other equivalently but some are interacting strongly stabilizing the molecule and deserving to be linked to each other by chemical bonds while the remaining pairs are just weakly interacting; these secondary interactions may have a stabilizing or destabilizing nature. Although this idea works well for diatomics and many polyatomic molecules, there are always examples of interactions that are hard to be classified as strong or weak stabilizing interactions thus demolishing a clear cut borderline between these two classes. This "gray" borderline, which is also the source of many controversies in the history of chemistry [13], was the main motivation of Bader's insistence that chemical bond is intrinsically a floppy concept that lacks a clear cut definition. However, accepting that in current situation the concept of chemical bond is inseparable from chemistry, how one may infer the presence of a chemical bond from the QTAIM analysis? Since, as also emphasized previously, there is no unique indictor of chemical bond in the QTAIM's toolbox, one needs to reemploy a *synthetic* approach. Ideally, in such synthetic approach, quantitative measures of the interaction strength of atoms in molecules with each other are the primary source of information as also advocated recently [14,15] however, any *basin property* of atoms in molecules, apart from the usual topological indexes, that shed light on the *communication* mode of atoms in molecules may contain valuable source of information. This synthetic approach that was introduced systematically in characterization of hydrogen and dihydrogen as well as agnostic bonds by Popelier [16-19] also employed by Farrugia and workers [2], using the delocalization index [3] as well as the source function [20] as typical basin properties, to infer that there are chemical bonds between iron atom and terminal carbons in $(CO)_3$Fe-TMM complex. In this regard, Table 1 presents the delocalization index between iron and carbon atoms at the four geometries depicted in Figure 2; manifestly, in contrast to the observed variation of MG, the delocalization index changes smoothly with no catastrophic change. At first step, it is evident from this table that the average number of electrons delocalized between iron atom and terminal carbons is always larger than the average number of electrons delocalized between central carbon and metal atom hinting that seeking chemical bonds between iron and terminal carbons is not illegitimate. Further considerations based on the QTAIM analysis [2], which shape a synthetic approach, confirm that assuming three chemical bonds between iron and terminal carbons is a legitimate choice that is also confirmed using the MO and EDA analysis [1].

To draw a more general conclusion for future studies one may ask a key question: Does the synthetic approach within the context of the QTAIM is capable of unraveling chemical bonds unambiguously? The answer is an explicit "no". Like the other theoretical approaches a synthetic approach within the context of the QTAIM needs a list of criteria, checking various topological and basin properties [19], which are used to infer the presence of a chemical bond between two atoms. Using such list as a guide, there will be always "gray" examples, stick with the border between bond and no-bond, and hard to be classified. On the other hand, such list, by its very nature, is not unique and some of its criteria may be replaced with other reasonable bonding criteria disclosing *rival* lists, a new source of ambiguity for gray examples. Obviously, this reasoning points to the fact that the presence of a chemical bond, though inferred from theoretical or experimental data, has always a *subjective* element that resists consolidation, a fact praised by some whereas criticized by others [6,21].

### 3. Conclusion

Both MG and BP/AIL, by their nature, are responding abruptly to nuclear excursions; some BP/AIL may appear/disappear at certain geometries thus their variations are not always continuous in regard to nuclear vibrations [3]. This fact, in itself, is against the usual assumed independence of models describing chemical bonds/chemical structures of nuclear vibrations thus against any proposal that equates BP and chemical bonds. Accordingly, by insisting on the equivalence of BP/AIL as descriptors of chemical bonds one is faced to accept one of these two unpleasant situations: 1) restricting, albeit *artificially*, the QTAIM analysis only to equilibrium geometries in contrast to the plain fact that molecules are not static objects or 2) incorporating MG of non-equilibrium geometries in the QTAIM analysis and claiming that chemical bonds of a molecule are formed and disrupt just by nuclear vibrations without any external interferences. None of these options seems to be legitimate; nuclear vibrational dynamics has been recently incorporated explicitly into the QTAIM analysis within context of the newly proposed multi-component QTAIM (MC-QTAIM) [22-26] thus its relevance to QTAIM analysis has been consolidated while probably there is a consensus among chemists that chemical bonds are *characteristics* of a molecule varying just by chemical transformations/reactions but not the internal dynamic of a molecule. It seems legitimate to generalize this conclusion and conclude that concept of MG is *not* a redefinition of chemical structure; a chemical structure is a *code* used to codify chemical (bonding and reactivity) as well as physical (spectroscopic and energetic) information of a molecule while a MG is a topological code revealing the topography of the one-electron density. Though in many examples the mathematical graphs of a MG and a chemical structure of an isolated molecule look quite similar, the methodologies behind their construction are apart thus ruin a universal correspondence of these two. Based on many computational QTAIM studies, one may claim that in most considered cases chemists draw bonds between atoms that share an IAS at the equilibrium geometry, the neighboring atoms, or share an IAS at more accessible non-equilibrium geometries, the passionate neighbors. However there is no rigorous reasoning to dismiss the possibility of attributing chemical bonds to atoms that are neighbors neither at equilibrium nor at more accessible non-equilibrium geometries.

One may sum up and conclude that the QTAIM is a theory of atoms in molecules and their properties relying basically on the tradition of experimental regularization of thermodynamic and spectroscopic data based on contributions of functional groups/atoms [27]. The data emerging from the QTAIM analysis are also useful to decipher the nature of chemical bonding, as usually used by most computational chemists, however, as emphasized also in the previous subsection, this must be done with caution to hesitate superficial conclusions and irrelevant controversies as some have been cited in ref. 28 of the original paper. Accordingly, *the QTAIM invents its own language and the emerging basic concepts are not always translatable in a one-to-one style into the orthodox concepts of the chemical language*.

**Acknowledgments**

The author is grateful to Cina Foroutan-Nejad and Masume Gharabaghi for their helpful suggestions on a previous draft of this paper and technical assistance.

References

1. Mousavi M, Frenking G (2013) J Organomet Chem doi: 10.1016 /j.jorganchem. 2013.03.047.
2. Farrugia LJ, Evans C, Tegel M (2006) J Phys Chem A 110: 7952
3. Bader RFW (1990) Atoms in Molecules: A Quantum Theory. Oxford University Press, Oxford
4. Bader RFW (2009) J Phys Chem A 113: 10391
5. Bone RGA, Bader RFW (1996) J Phys Chem 100: 10892
6. Bader RFW (2010) J Phys Chem A 114: 7431
7. Shahbazian Sh, Sadjadi A (2007) J Mol Struct (Theochem) 822: 116
8. Shahbazian Sh, Alizadeh Sh (2008) J Phys Chem A 112: 10365
9. Biegler-König F (2000) J Comput Chem 21: 1040
10. Biegler-König F, Schönbohm J, Bayles D (2001) J Comput Chem 22: 545
11. Biegler-König F, Schönbohm J (2002) J Comput Chem 23: 1489
12. Pauling L (1960) The Nature of Chemical Bond. Oxford & IBH Publishing Co.: New Delhi, third edition
13. Nye MJ (1993) From Chemical Philosophy to Theoretical Chemistry. University of California Press: Berkeley
14. García-Revilla M, Francisco E, Popelier PLA, Martin Pendas A (2013) ChemPhysChem 14: 1211
15. Dem'yanov P, Polestshuk P (2012) Chem Eur J 18: 4982
16. koch U, Popelier PLA (1995) J Phys Chem 99: 9747
17. Popelier PLA (1998) J Phys Chem 102: 1873
18. Popelier PLA, Logothetis G (1998) J Organometal Chem 555: 101
19. Popelier PLA (2000) Atoms in Molecules an Introduction. Pearson, London
20. Bader RFW, Gatti C (1998) Chem Phys Lett 287: 233
21. Shahbazian Sh, Zahedi M (2007) Found Chem 9: 85
22. Goli M, Shahbazian Sh (2011) Theor Chem Acc 129: 235
23. Goli M, Shahbazian Sh (2012) Theor Chem Acc 131: 1208
24. Goli M, Shahbazian Sh (2013) Theor Chem Acc 132: 1362
25. Goli M, Shahbazian Sh (2013) Theor Chem Acc 132: 1365

26. Shahbazian Sh (2013) Found Chem doi: 10.1007/s10698-012-9170-0
27. Matta CF, Bader RFW (2006) J Phys Chem A 110: 6365

Figure Captions:

Figure 1. The MG of (CO)$_3$Fe-TMM complex  a) at the equilibrium geometry, Fe-C$_{terminal}$ = 2.141, Fe-C$_{centeral}$ = 1.955, C$_{terminal}$-C$_{centeral}$ = 1.436 b) Fe-C$_{terminal}$ = 2.028, Fe-C$_{centeral}$ = 1.886, C$_{terminal}$-C$_{centeral}$ = 1.453 c) Fe-C$_{terminal}$ = 1.91, Fe-C$_{centeral}$ = 1.807, C$_{terminal}$-C$_{centeral}$ = 1.477 d) Fe-C$_{terminal}$ = 1.86, Fe-C$_{centeral}$ = 1.777, C$_{terminal}$-C$_{centeral}$ = 1.488.  The carbon, oxygen, iron and hydrogen nuclei are colored as larger black, red, white and gray spheres, respectively, while the smaller red and yellow spheres are bond and ring critical points, respectively. The BP and AIL are also depicted for each linked pair of atoms. All geometrical values are in angstroms.

Figure 2. The MG of (CO)$_3$Fe-TMM complex  a) at the equilibrium geometry, Fe-C$_{terminal}$ = 2.141, Fe-C$_{centeral}$ = 1.955, C$_{terminal}$-C$_{centeral}$ = 1.436 b) Fe-C$_{terminal}$ = 2.100, Fe-C$_{centeral}$ = 1.955, C$_{terminal}$-C$_{centeral}$ = 1.441 c) Fe-C$_{terminal}$ = 2.055, Fe-C$_{centeral}$ = 1.955, C$_{terminal}$-C$_{centeral}$ = 1.448 d) Fe-C$_{terminal}$ = 2.005, Fe-C$_{centeral}$ = 1.955, C$_{terminal}$-C$_{centeral}$ = 1.457.  The carbon, oxygen, iron and hydrogen nuclei are colored as larger black, red, white and gray spheres, respectively, while the smaller red and yellow spheres are bond and ring critical points, respectively. The BP and AIL are also depicted for each linked pair of atoms. All geometrical values are in angstroms (the target terminal carbon in dataset is the one nearest to iron nuclei).

Table 1- The computed delocalization index for certain pair of atoms at the equilibrium and three non-equilibrium geometries of $(CO)_3$Fe-TMM complex depicted in Figure 2.

| Configuration | $C_{triminal}$-$C_{center}$ | Fe-$C_{centeral}$ | Fe-$C_{terminal}$ |
|---|---|---|---|
| *Figure-2a* | 1.21 | 0.40 | 0.58 |
| *Figure-2b* | 1.19 | 0.39 | 0.60 |
| *Figure-2c* | 1.17 | 0.39 | 0.62 |
| *Figure-2d* | 1.16 | 0.39 | 0.65 |

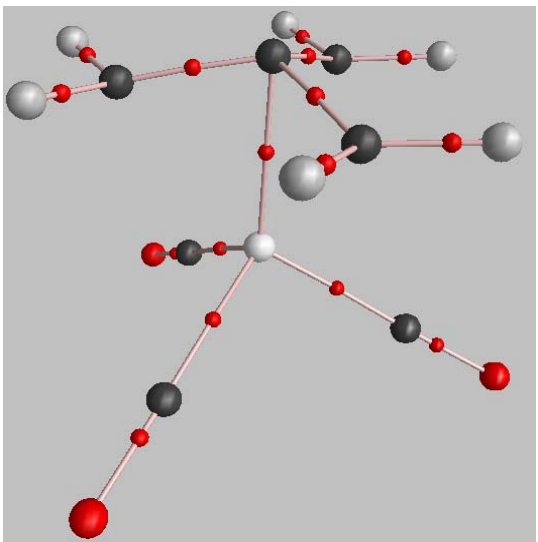 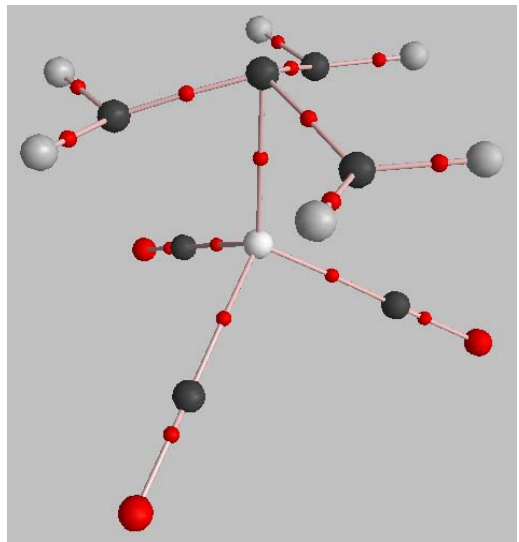

(a)        (b)

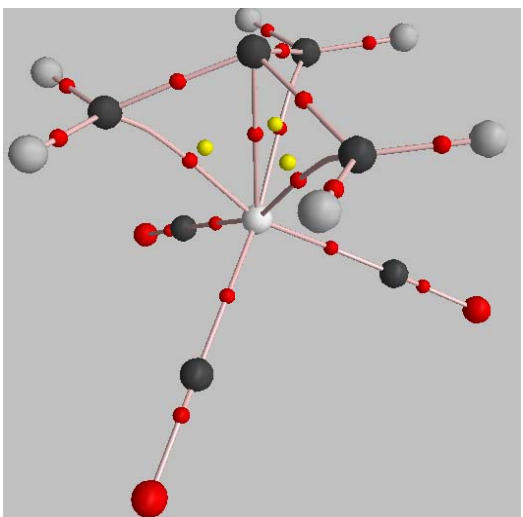 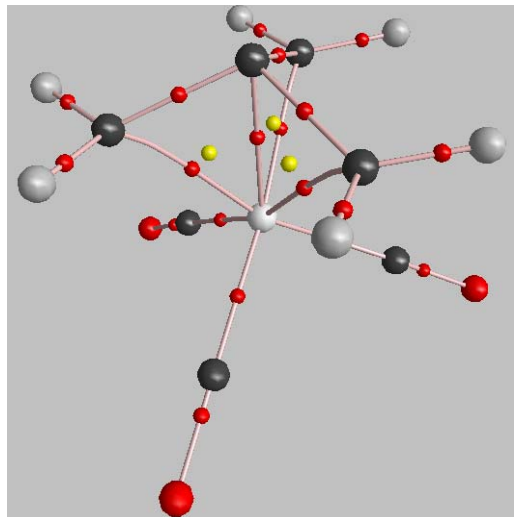

(c)        (d)

Figure 1

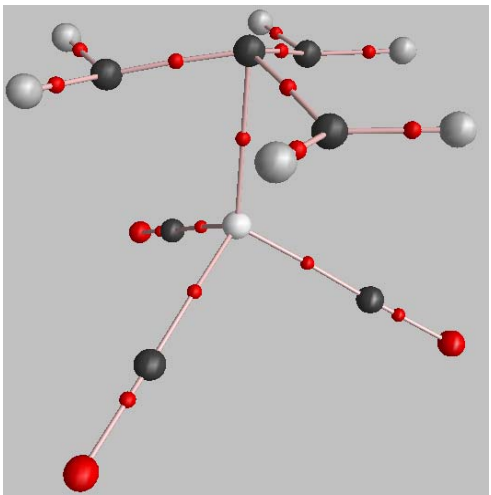 (a)
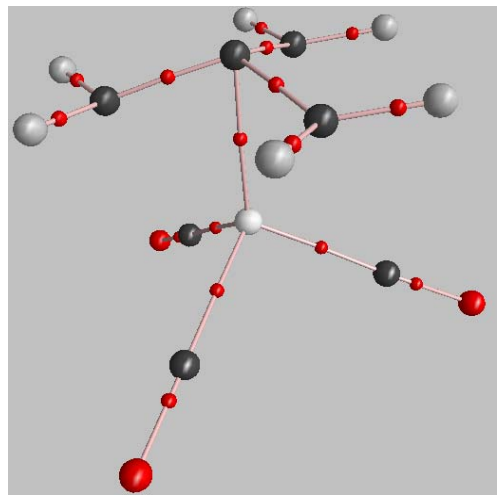 (b)
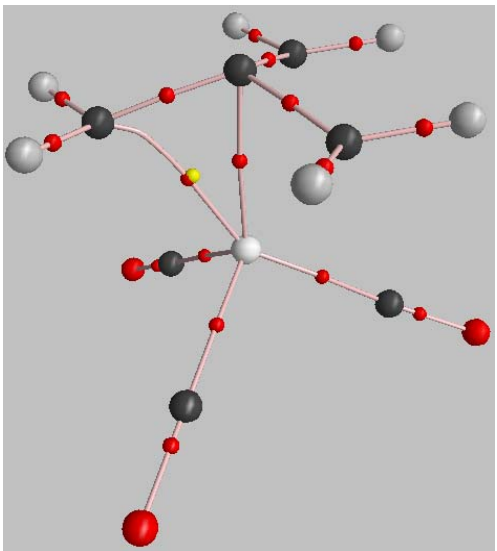 (c)
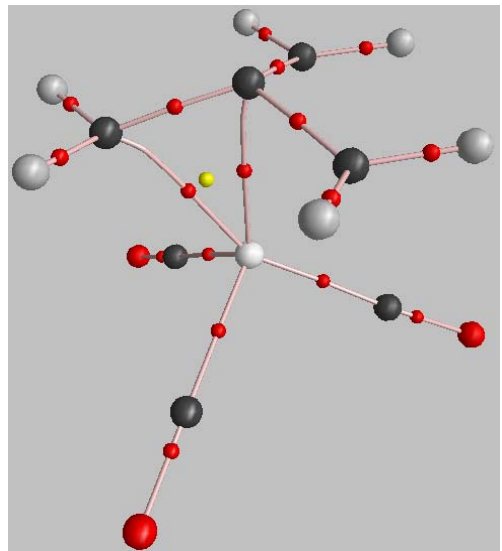 (d)

Figure 2